# A Conceptual Technology-Dependent Framework of Ternary Quantum Gates

Ali Al-Bayaty
Department of Electrical and Computer Engineering
Portland State University
Portland, OR 97201, USA
albayaty@pdx.edu

***Abstract–***This paper introduces a conceptual framework of technology-dependent ternary quantum gates that could be implemented and fabricated into future superconducting and photonic quantum systems for operating 3-valued quantum bits (qutrits). The "technology-dependent" means that such ternary quantum gates are on-purpose designed analogy to the contemporary binary quantum gates. Conceptually, the final built technology-dependent one-, two-, and three-qutrit gates are Chrestenson, Chrestenson$^{\dagger}$, $Z_3$, $Z_3^{\dagger}$, 01, 02, 12, +1, +2 (including their corresponding controlled gates), a non-phase relative SWAP gate, and a cost-effective Toffoli gate, which is a generic ternary Galois Field (GF3) multiplication and addition circuit.

## 1. Introduction

The binary quantum computing has been thoroughly discussed in various literature [1–5], with its applicable approaches and search algorithms [6–11] to solve many sets of problems, which are constructed as Boolean or phase oracles [8–11]. In binary quantum computing, the utilized quantum information is formulated using the 2-valued quantum bits (qubits), where a qubit (without any global phase) can hold (i) the basis binary states of $|0\rangle$ and $|1\rangle$, or (ii) the superimposed binary states of $|0 \pm 1\rangle$ and $|0 \pm i1\rangle$ when the binary superposition gates of Hadamard (H) and $\sqrt{\text{X}}$ (or $\sqrt{\text{X}}^{\dagger}$) are applied, respectively, where $i = e^{\pm\pi}$ radians. Notice that when utilizing $n$ qubits, the overall quantum space computation is restricted to $2^n$. For this reason, the ternary quantum computing is introduced [12–16] to increase the overall quantum space computation to $3^m$, using $m$ 3-valued quantum information, which is termed "qutrits" [12–16]. Notice that a qutrit (without any global phase) can hold (i) the basis ternary states of $|0\rangle$, $|1\rangle$, and $|2\rangle$, or (ii) the superimposed ternary states of $|0 \pm 1 \pm 2\rangle$, $|0 \pm \omega 1 \pm \omega^2 2\rangle$, and $|0 \pm \omega^2 1 \pm \omega 2\rangle$ when the ternary Chrestenson (CH or $C_3$) superposition gate is applied [12,17–20], where $\omega = e^{\frac{\pm 2\pi}{3}}$ radians.

Technically, in all binary quantum computing systems, the H gate is a non-native (non-supported) gate. For this reason, the H gate is constructed using a set of native gates for a specific technology-dependent binary quantum system. For instance, in the binary IBM Kingston quantum computer [21,22], the H gate is the sequence of three native gates {RZ(+π), $\sqrt{X}$, RZ(+π)}, where the native gates of this quantum computer are the Identity (I), Pauli-X (X), half-rotational X ($\sqrt{X}$), rotational X (RX), rotational Pauli-Z (RZ), rotational Z-Z (RZZ), and controlled-Z (CZ) gates.

The goal of this paper is to conceptually introduce a framework of technology-dependent ternary quantum gates that could be implemented and fabricated into future superconducting and photonic quantum systems. These ternary quantum gates are on-purpose designed analogies to the contemporary binary quantum gates, as one-, two-, and three-qutrit gates as follows.

1. Chrestenson (CH): Ternary superposition gate analogy to the binary H superposition gate.
2. Controlled-Chrestenson (CCH): Two-qutrit controlled superposition gate.
3. Chrestenson$^\dagger$ (CH$^\dagger$): The inverse superposition gate of the ternary CH gate.
4. Controlled-Chrestenson$^\dagger$ (CCH$^\dagger$): Two-qutrit controlled inverse superposition gate.
5. $Z_3$: Ternary phase gate analogy to the binary Z phase gate.
6. Controlled-$Z_3$ ($CZ_3$): Two-qutrit controlled phase gate.
7. $Z_3^\dagger$: The inverse phase gate of the ternary $Z_3$ gate.
8. Controlled-$Z_3^\dagger$ ($CZ_3^\dagger$): Two-qutrit controlled inverse phase gate.
9. 01, 02, and 12: Ternary permutative gates for switching the ternary state of a qutrit.
10. Controlled-01 (C01), controlled-02 (C02), and controlled-12 (C12): Two-qutrit gates.
11. +1 and +2: Ternary shift ternary gates for incrementing the ternary state of a qutrit.
12. Controlled-+1 (C+1) and controlled-+2 (C+2): Two-qutrit gates.
13. SWAP: Two-qutrit non-phase relative swapping gate.
14. Toffoli: Three-qutrit gate, a ternary Galois Field (GF3) multiplication and addition circuit.

## 2. Methods

In ternary quantum computing, the well-known one-qutrit superposition gate is the Chrestenson (CH or $C_3$) gate [12,17–20] and the one-qutrit phase gate is the $Z_3$ gate [12,23,24], as expressed in Eq. (1) and Eq. (2), respectively, where $\omega = e^{\frac{\pm 2\pi}{3}}$ radians, $\omega = -\,\omega^2$, and $\omega^2 = -\,\omega$. The $Z_3$ gate applies a no-phase, $\omega$ phase, or $\omega^2$ phase to the $|0\rangle$, $|1\rangle$, or $|2\rangle$ state of a qutrit, respectively.

Mathematically, the $Z_3^\dagger$ gate can be directly constructed using two $Z_3$ gates, as expressed in Eq. (3). Notice that the $Z_3^\dagger$ gate applies a no-phase, $\omega^2$ phase, or $\omega$ phase to the $|0\rangle$, $|1\rangle$, or $|2\rangle$ state of a qutrit, respectively. In general, the CH and $Z_3$ gates are reset to one Identity ($I_3$) gate for a number of sequential repetitions, as stated in Eq. (4) and Eq. (5), respectively. Notice that, using the periodical geometry technique, $1+\omega+\omega^2 = 0$, $\omega^3 = 1$, $\omega^4 = \omega$, $\omega^5 = \omega^2$, and so on.

$$\mathrm{CH} = \frac{1}{\sqrt{3}}\begin{bmatrix} 1 & 1 & 1 \\ 1 & \omega & \omega^2 \\ 1 & \omega^2 & \omega \end{bmatrix} \tag{1}$$

$$\mathrm{Z}_3 = \begin{bmatrix} 1 & 0 & 0 \\ 0 & \omega & 0 \\ 0 & 0 & \omega^2 \end{bmatrix} \tag{2}$$

$$\mathrm{Z}_3{}^\dagger = \mathrm{Z}_3 \cdot \mathrm{Z}_3 = \begin{bmatrix} 1 & 0 & 0 \\ 0 & \omega^2 & 0 \\ 0 & 0 & \omega \end{bmatrix} \tag{3}$$

$$\mathrm{CH} \cdot \mathrm{CH} \cdot \mathrm{CH} \cdot \mathrm{CH} = \mathrm{I}_3 = \begin{bmatrix} 1 & 0 & 0 \\ 0 & 1 & 0 \\ 0 & 0 & 1 \end{bmatrix} \tag{4}$$

$$\mathrm{Z}_3 \cdot \mathrm{Z}_3 \cdot \mathrm{Z}_3 = \mathrm{Z}_3 \cdot \mathrm{Z}_3{}^\dagger = \mathrm{Z}_3{}^\dagger \cdot \mathrm{Z}_3 = \mathrm{I}_3 = \begin{bmatrix} 1 & 0 & 0 \\ 0 & 1 & 0 \\ 0 & 0 & 1 \end{bmatrix} \tag{5}$$

### 2.1 Ternary Gates Design Postulations

Our mythological approach to conceptually designing one-qutrit and two-qutrit quantum gates is categorized into three postulations, by fundamentally assuming the pre-implemented technology-dependent ternary native gates for a specific ternary quantum computer. These three postulations are basically based on the construction of binary non-native gates using a set of binary native gates (as the pre-implemented technology-dependent gates) for a specific binary quantum computer.

For instance, in the binary IBM Kingston quantum computer, the non-native H superposition gate is built in the sequence of native gates {RZ(+π), $\sqrt{X}$, RZ(+π)}. Notice that the $\sqrt{X}$ gate is also considered a binary superposition gate, but on a different axis of the Bloch sphere, as shown in Fig. 1. That means the H gate is constructed from one native superposition gate ($\sqrt{X}$) and two native rotational gates (RZ). Hence, the H gate is a (non-implemented) technology-independent superposition gate, while the $\sqrt{X}$ gate is a (pre-implemented) technology-dependent superposition gate. Similarly, the CH gate can be constructed using one pre-implemented technology-dependent ternary superposition gate (as a postulation) and a number of ternary rotational gates ($Z_3$ and $Z_3^\dagger$).

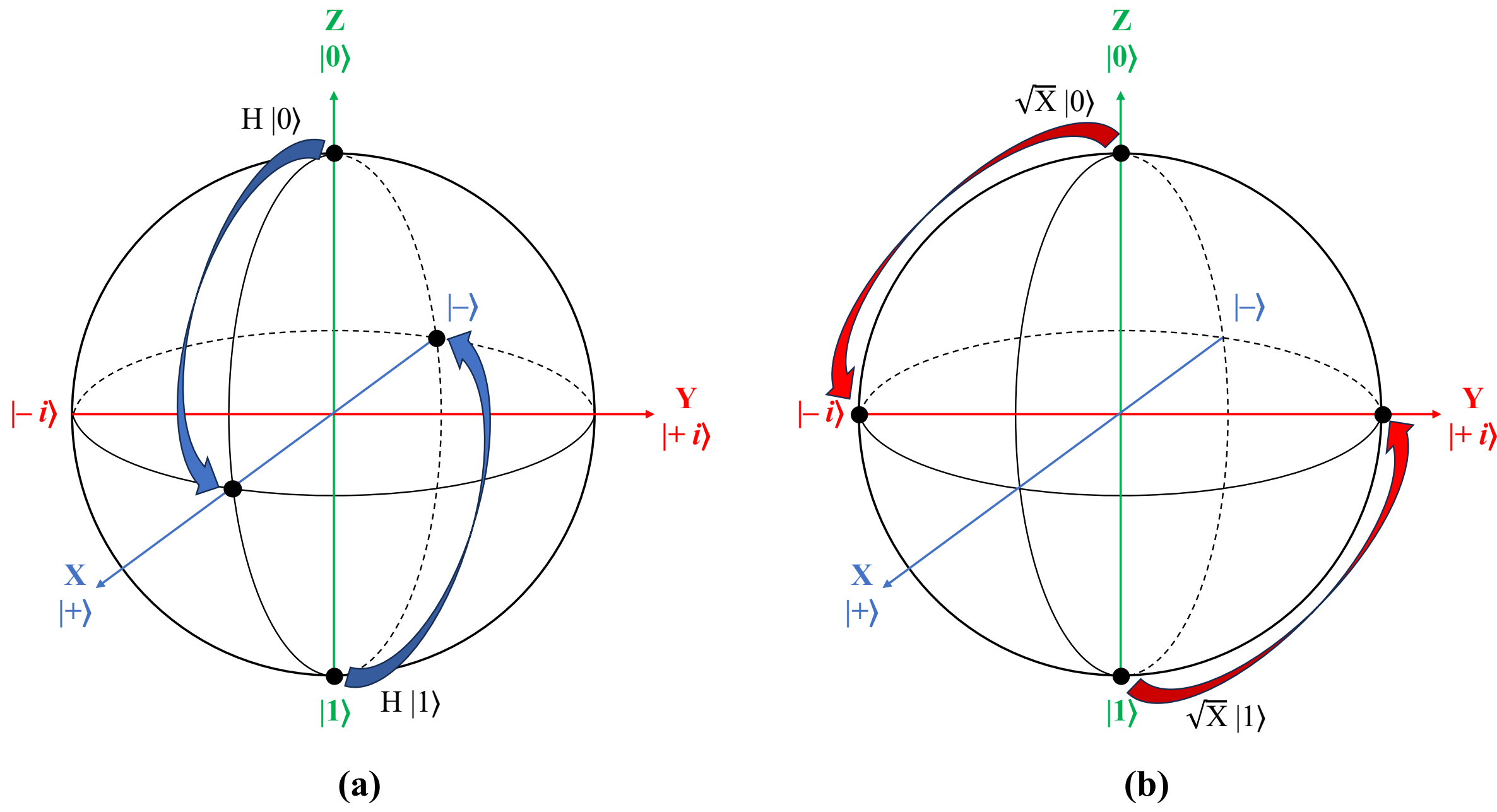

**Figure 1.** The Bloch sphere is a three-dimensional geometrical sphere: **(a)** a Hadamard (H) gate creates superimposed states on the X-axis, and **(b)** a $\sqrt{\text{X}}$ gate creates superimposed states on the Y-axis [25].

**Postulation I:** *Assume the ternary $Z_3$ gate and ternary superposition gate ($TSG_I$), as expressed in Eq. (6), are pre-implemented technology-dependent native gates for a specific ternary quantum computer. Such that, the Chrestenson (CH) superposition gate is then directly constructed, as stated in Eq. (7). Mathematically, the $TSG_I$ gate can be directly calculated using the reverse matrices multiplication, since the CH and $Z_3$ are well-known ternary gates.*

$$\text{TSG}_{\text{I}} = \frac{1}{\sqrt{3}}\begin{bmatrix} 1 & \omega^2 & \omega \\ \omega^2 & \omega^2 & \omega^2 \\ \omega & \omega^2 & 1 \end{bmatrix} \tag{6}$$

$$\text{CH} = \text{Z}_3 \cdot \text{TSG}_{\text{I}} \cdot \text{Z}_3 = \frac{1}{\sqrt{3}}\begin{bmatrix} 1 & 1 & 1 \\ 1 & \omega & \omega^2 \\ 1 & \omega^2 & \omega \end{bmatrix} \tag{7}$$

□

**Postulation II:** *Assume the ternary $Z_3$ gate, ternary $Z_3^\dagger$ gate, and ternary superposition gate ($TSG_{II}$), as expressed in Eq. (8), are pre-implemented technology-dependent native gates for a specific ternary quantum computer. Such that, the Chrestenson (CH) superposition gate is then directly constructed, as stated in Eq. (9). Mathematically, the $TSG_{II}$ gate can be directly calculated using the reverse matrices multiplication, since the CH and $Z_3$ are well-known ternary gates.*

$$\mathrm{TSG_{II}} = \frac{1}{\sqrt{3}}\begin{bmatrix} 1 & \omega^2 & \omega \\ \omega & \omega & \omega \\ \omega^2 & 1 & \omega \end{bmatrix} \quad (8)$$

$$\mathrm{CH} = \mathrm{Z_3} \cdot \mathrm{TSG_{II}} \cdot \mathrm{Z_3}^{\dagger} = \frac{1}{\sqrt{3}}\begin{bmatrix} 1 & 1 & 1 \\ 1 & \omega & \omega^2 \\ 1 & \omega^2 & \omega \end{bmatrix} \quad (9)$$

□

**Postulation III:** *Assume the ternary $Z_3$ gate and ternary superposition gate ($TSG_{III}$), as expressed in Eq. (10), are pre-implemented technology-dependent native gates for a specific ternary quantum computer. Such that, the Chrestenson (CH) superposition gate is then directly constructed, as stated in Eq. (11). Mathematically, the $TSG_{III}$ gate can be directly calculated using the reverse matrices multiplication, since the CH and $Z_3$ are well-known ternary gates.*

$$\mathrm{TSG_{III}} = \frac{1}{\sqrt{3}}\begin{bmatrix} 1 & 1 & 1 \\ \omega^2 & 1 & \omega \\ \omega & 1 & \omega^2 \end{bmatrix} \quad (10)$$

$$\mathrm{CH} = \mathrm{TSG_{III}} \cdot \mathrm{Z_3} = \frac{1}{\sqrt{3}}\begin{bmatrix} 1 & 1 & 1 \\ 1 & \omega & \omega^2 \\ 1 & \omega^2 & \omega \end{bmatrix} \quad (11)$$

□

Notice that, for all aforementioned three postulations, the ternary $Z_3$ gate always has to be a pre-implemented technology-dependent native gate for a specific ternary quantum computer. Depending on the pre-implemented technology-dependent native TSG (with $Z_3$ and/or $Z_3^{\dagger}$) gates, identical ternary CH superposition gates are successfully constructed.

Moreover, using any of these three postulations, the ternary $CH^{\dagger}$ superposition gate can also be constructed using a set of $Z_3$, $Z_3^{\dagger}$, and $TSG^{\dagger}$ gates. For instance, using Postulation II, the $CH^{\dagger}$ gate is constructed using the $TSG_{II}^{\dagger}$, as stated in Eq. (12), $Z_3$, and $Z_3^{\dagger}$ gates for the final inverse CH gate construction expressed in Eq. (13). Notice that $TSG_{II} \cdot TSG_{II}^{\dagger} = I_3$.

$$\mathrm{TSG_{II}}^{\dagger} = \frac{1}{\sqrt{3}}\begin{bmatrix} 1 & \omega^2 & \omega \\ \omega & \omega^2 & 1 \\ \omega^2 & \omega^2 & \omega^2 \end{bmatrix} \quad (12)$$

$$\mathrm{CH}^{\dagger} = \mathrm{Z_3} \cdot \mathrm{TSG_{II}}^{\dagger} \cdot \mathrm{Z_3}^{\dagger} = \frac{1}{\sqrt{3}}\begin{bmatrix} 1 & 1 & 1 \\ 1 & \omega^2 & \omega \\ 1 & \omega & \omega^2 \end{bmatrix} \quad (13)$$

Notably, the advantage of constructing the $CH^\dagger$ gate in ternary quantum computing is that the CH gate can be immediately reset using only two gates in sequence, i.e., $CH \cdot CH^\dagger = I_3$, without using the expensive four-gate resetting technique, as stated in Eq. (4) above. On the one hand, using Postulation II, the four-gate resetting technique of CH gate is a cost-expensive approach, which utilizes four $Z_3$, four $Z_3^\dagger$, and four $TSG_{II}$ gates, i.e., a total of 12 ternary gates, as stated in Eq. (14). On the other hand, using Postulation II, the two-gate resetting technique of CH gate is a cost-effective approach, which utilizes two $Z_3$, two $Z_3^\dagger$, one $TSG_{II}$, and one $TSG_{II}^\dagger$ gates, i.e., a total of six ternary gates, as stated in Eq. (15). Notice that $TSG_{II} \cdot TSG_{II} \cdot TSG_{II} \cdot TSG_{II} = I_3$.

$$\begin{aligned} CH \cdot CH \cdot CH \cdot CH &= (Z_3 \cdot TSG_{II} \cdot Z_3^\dagger) \cdot (Z_3 \cdot TSG_{II} \cdot Z_3^\dagger) \cdot (Z_3 \cdot TSG_{II} \cdot Z_3^\dagger) \cdot (Z_3 \cdot TSG_{II} \cdot Z_3^\dagger) \\ &= Z_3 \cdot TSG_{II} \cdot I_3 \cdot TSG_{II} \cdot I_3 \cdot TSG_{II} \cdot I_3 \cdot TSG_{II} \cdot Z_3^\dagger = Z_3 \cdot I_3 \cdot Z_3^\dagger = I_3 \end{aligned} \tag{14}$$

$$\begin{aligned} CH \cdot CH^\dagger &= (Z_3 \cdot TSG_{II} \cdot Z_3^\dagger) \cdot (Z_3 \cdot TSG_{II}^\dagger \cdot Z_3^\dagger) \\ &= Z_3 \cdot TSG_{II} \cdot I_3 \cdot TSG_{II}^\dagger \cdot Z_3^\dagger = Z_3 \cdot I_3 \cdot Z_3^\dagger = I_3 \end{aligned} \tag{15}$$

## 2.2 Ternary Permutative and Shift Gates

Because all the aforementioned three postulations construct identical CH gates by mainly using the $Z_3$ gates, the well-known ternary permutative (01, 02, and 12) gates [12,18] can be directly generated using only CH and $Z_3$ gates, as stated in Eq. (16), Eq. (17), and Eq. (18), respectively. Similarly, the well-known ternary shift (+1 and +2) gates [12,14,18] can be directly generated using only these ternary permutative gates, as stated in Eq. (19) and Eq. (20), respectively. From Eq. (16) to Eq. (20), all ternary gates are ordered based on how they appear visually (sequentially arranged) on-the-wire for a qutrit in a ternary quantum circuit. Notice that $+1 \cdot +2 = +2 \cdot +1 = I_3$.

$$01 = CH \cdot Z_3^\dagger \cdot CH = CH \cdot Z_3 \cdot Z_3 \cdot CH = 10 = \begin{bmatrix} 0 & 1 & 0 \\ 1 & 0 & 0 \\ 0 & 0 & 1 \end{bmatrix} \tag{16}$$

$$02 = CH \cdot Z_3 \cdot CH = 20 = \begin{bmatrix} 0 & 0 & 1 \\ 0 & 1 & 0 \\ 1 & 0 & 0 \end{bmatrix} = \text{ternary NOT gate} \tag{17}$$

$$12 = CH \cdot CH = 21 = \begin{bmatrix} 1 & 0 & 0 \\ 0 & 0 & 1 \\ 0 & 1 & 0 \end{bmatrix} \tag{18}$$

$$+1 = 01 \cdot 02 = 12 \cdot 01 = 02 \cdot 12 = +2 \cdot +2 = -2 = \begin{bmatrix} 0 & 0 & 1 \\ 1 & 0 & 0 \\ 0 & 1 & 0 \end{bmatrix} \quad (19)$$

$$+2 = 02 \cdot 01 = 01 \cdot 12 = 12 \cdot 02 = +1 \cdot +1 = -1 = \begin{bmatrix} 0 & 1 & 0 \\ 0 & 0 & 1 \\ 1 & 0 & 0 \end{bmatrix} \quad (20)$$

## 2.3 Ternary Controlled Gates

By assuming controlled versions (two-qutrit) of the aforementioned three postulations with their pre-implemented technology-dependent native gates for a specific ternary quantum computer, i.e., the controlled-$Z_3$ ($CZ_3$), controlled-$Z_3^\dagger$ ($CZ_3^\dagger$), controlled-$TSG_I$ ($CTSG_I$), controlled-$TSG_{II}$ ($CTSG_{II}$), controlled-$TSG_{II}^\dagger$ ($CTSG_{II}^\dagger$), and controlled-$TSG_{III}$ ($CTSG_{III}$) gates, various ternary controlled gates are immediately generated, i.e., the controlled-CH (CCH), controlled-$CH^\dagger$ ($CCH^\dagger$), controlled-permutative (C01, C02, and C12), and controlled-shift (C+1 and C+2) gates.

**Example 1.** Using Postulation II, the CCH is constructed as stated in Eq. (21) and illustrated in Fig. 2. Such that, in Fig. 2, when the control qutrit ($q_a$) is initially set to the $|2\rangle$ state, the $CTSG_{II}$ is activated and the CCH is constructed, as the CH gate for the target qutrit ($q_b$). Otherwise, when the $q_a$ is initially set to either the $|0\rangle$ or $|1\rangle$ state, the CCH is not constructed, because the $CTSG_{II}$ is not activated and the $Z_3$ cancels $Z_3^\dagger$, as stated in Eq. (5) above.

$$CCH = Z_3 \cdot CTSG_{II} \cdot Z_3^\dagger \quad (21)$$

**Figure 2.** A decomposed two-qutrit circuit of CCH, where $q_a$ is the control qutrit and $q_b$ is the target qutrit.

**Example 2.** Using Postulation II, the permutative controlled-12 (C12) gate is constructed as stated in Eq. (22) and shown in Fig. 3. Such that, in Fig. 3, when the control qutrit ($q_a$) is initially set to the $|2\rangle$ state, all two $CTSG_{II}$ are activated and the C12 is constructed, as the 12 gate for the target qutrit ($q_b$). Otherwise, when the $q_a$ is initially set to either the $|0\rangle$ or $|1\rangle$ state, the C12 is not constructed, because all two $CTSG_{II}$ are not activated and the $Z_3$ cancels $Z_3^\dagger$.

$$
\begin{aligned}
C12 = CCH \cdot CCH &= (Z_3 \cdot CTSG_{II} \cdot Z_3^{\dagger}) \cdot (Z_3 \cdot CTSG_{II} \cdot Z_3^{\dagger}) \\
&= Z_3 \cdot CTSG_{II} \cdot CTSG_{II} \cdot Z_3^{\dagger} \qquad (22)
\end{aligned}
$$

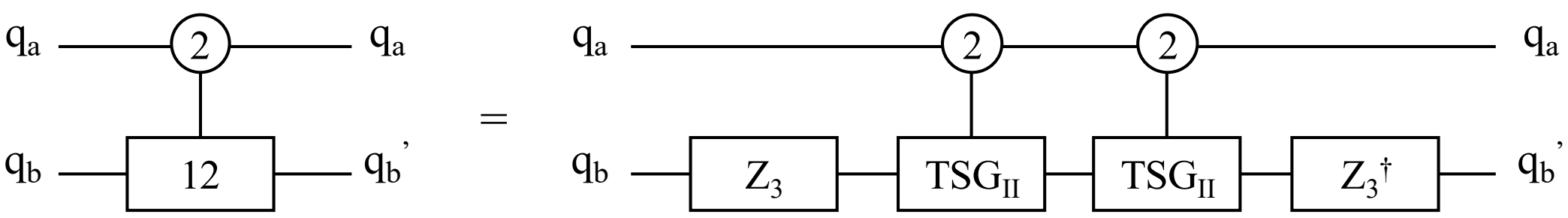

**Figure 3.** A decomposed two-qutrit circuit of C12, where $q_a$ is the control qutrit and $q_b$ is the target qutrit.

**Example 3.** Using Postulation II, the shift controlled-+1 (C+1) gate is constructed as demonstrated in Eq. (23) and shown in Fig. 4. Such that, in Fig. 4, when the control qutrit ($q_a$) is initially set to the $|2\rangle$ state, all four $CTSG_{II}$ are activated and the C+1 is constructed, as the +1 gate for the target qutrit ($q_b$). Otherwise, when the $q_a$ is initially set to either the $|0\rangle$ or $|1\rangle$ state, the C+1 is not constructed, because all four $CTSG_{II}$ are not activated and the two $Z_3$ cancel the two $Z_3^{\dagger}$.

$$
\begin{aligned}
C{+}1 = C01 \cdot C02 &= (CCH \cdot Z_3^{\dagger} \cdot CCH) \cdot (CCH \cdot Z_3 \cdot CCH) \\
&= Z_3 \cdot CTSG_{II} \cdot Z_3^{\dagger} \cdot Z_3^{\dagger} \cdot Z_3 \cdot CTSG_{II} \cdot Z_3^{\dagger} \cdot Z_3 \cdot CTSG_{II} \cdot Z_3^{\dagger} \cdot Z_3 \cdot Z_3 \cdot CTSG_{II} \cdot Z_3^{\dagger} \\
&= Z_3 \cdot CTSG_{II} \cdot Z_3^{\dagger} \cdot CTSG_{II} \cdot CTSG_{II} \cdot Z_3 \cdot CTSG_{II} \cdot Z_3^{\dagger} \qquad (23)
\end{aligned}
$$

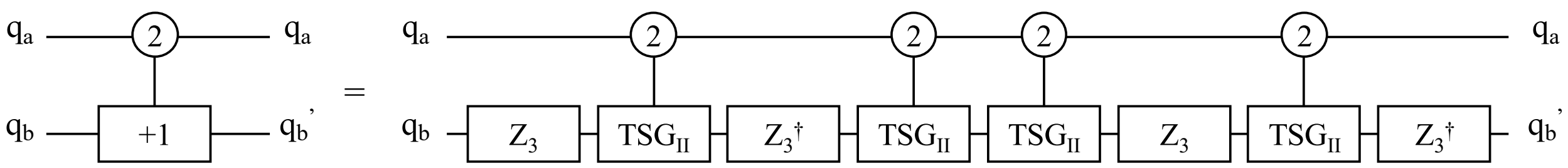

**Figure 4.** A decomposed two-qutrit circuit of C+1, where $q_a$ is the control qutrit and $q_b$ is the target qutrit.

## 3. Results and Discussion

Our introduced conceptual technology-dependent of three postulations for constructing one-qutrit and two-qutrit cost-effective ternary quantum gates, as one framework, is summarized as illustrated in Fig. 5.

Notably, the controlled-$Z_3$ ($CZ_3$) and controlled-$Z_3^{\dagger}$ ($CZ_3^{\dagger}$) gates are reserved for our future work in constructing various cost-effective ternary quantum gates, in the context of the final total number of the utilized one-qutrit and two-qutrit gates in a ternary quantum circuit, which can be defined as a ternary "quantum cost ($QC_3$)" metric.

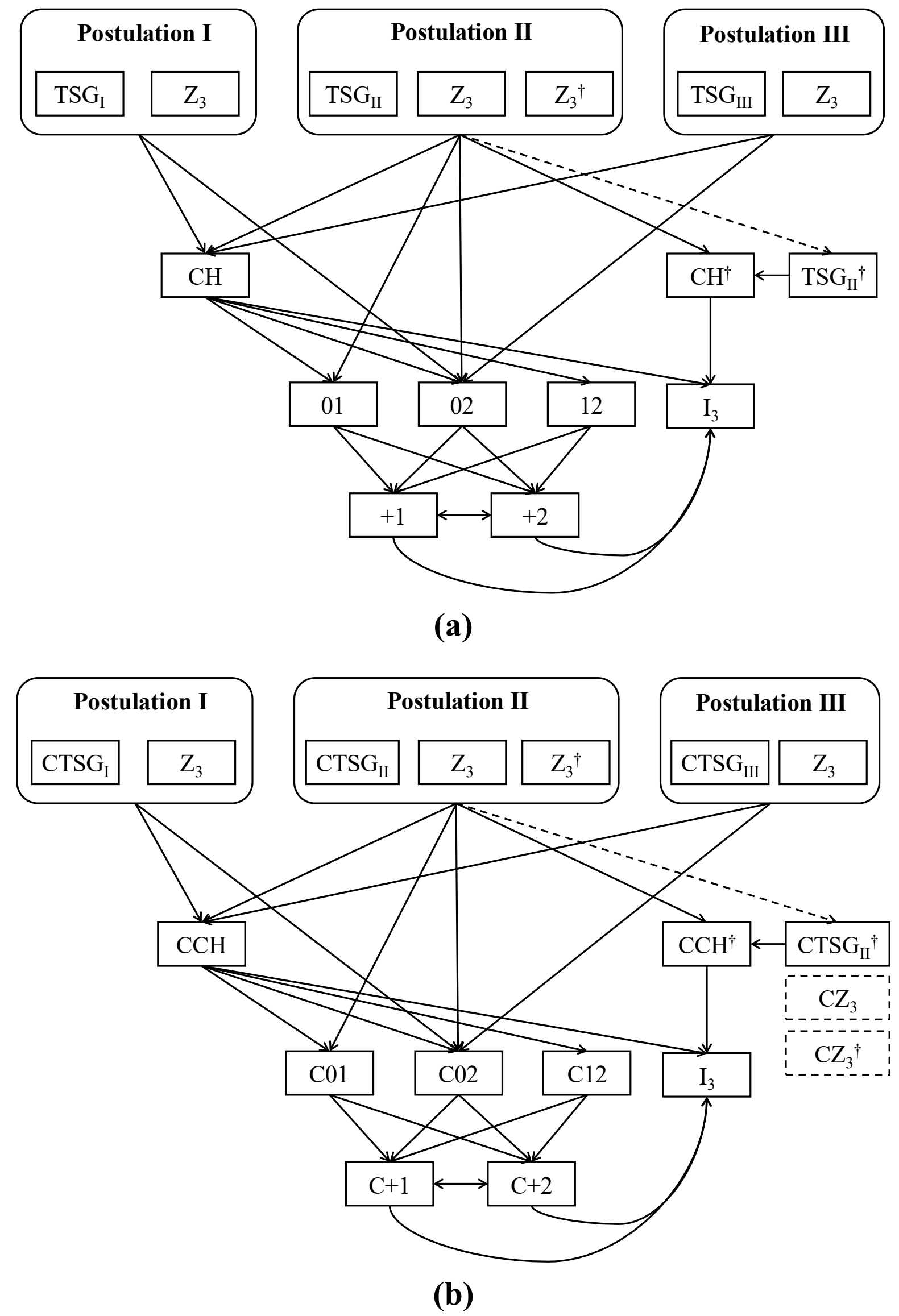

**Figure 5.** Our conceptual technology-dependent of three postulations (as one framework) for constructing: **(a)** one-qutrit cost-effective ternary quantum gates, and **(b)** two-qutrit cost-effective ternary quantum gates.

Additionally, based on (i) the aforementioned three postulations for one-qutrit gates, (ii) the presented two-qutrit controlled permutative and shift gates in the previous section, and (iii) our previous work in constructing a cost-effective binary *p*-SWAP gate [26,27], the quantum circuit of a non-phase relative ternary SWAP gate is conceptually proposed, as expressed in Eq. (24), stated in Table 1, and shown in Fig. 6, with the stage-by-stage analysis for the outcome states of two qutrits, as stated in Table 2. Notice that, in Fig. 6, the quantum cost ($QC_3$) of this ternary SWAP gate is nine ternary gates consisting of three one-qutrit gates and six two-qutrit gates.

$$|q_b\ q_a\rangle \rightarrow |q_a\ q_b\rangle;\ \forall\ q_a, q_b \in \{0, 1, 2\} \tag{24}$$

**Table 1.** The Marquand chart of our proposed non-phase relative ternary SWAP gate.

| $|q_b\ q_a\rangle$ \ $|q_a\ q_b\rangle$ | **$|00\rangle$** | **$|01\rangle$** | **$|02\rangle$** | **$|10\rangle$** | **$|11\rangle$** | **$|12\rangle$** | **$|20\rangle$** | **$|21\rangle$** | **$|22\rangle$** |
|---|---|---|---|---|---|---|---|---|---|
| **$|00\rangle$** | **1** | 0 | 0 | 0 | 0 | 0 | 0 | 0 | 0 |
| **$|01\rangle$** | 0 | 0 | 0 | **1** | 0 | 0 | 0 | 0 | 0 |
| **$|02\rangle$** | 0 | 0 | 0 | 0 | 0 | 0 | **1** | 0 | 0 |
| **$|10\rangle$** | 0 | **1** | 0 | 0 | 0 | 0 | 0 | 0 | 0 |
| **$|11\rangle$** | 0 | 0 | 0 | 0 | **1** | 0 | 0 | 0 | 0 |
| **$|12\rangle$** | 0 | 0 | 0 | 0 | 0 | 0 | 0 | **1** | 0 |
| **$|20\rangle$** | 0 | 0 | **1** | 0 | 0 | 0 | 0 | 0 | 0 |
| **$|21\rangle$** | 0 | 0 | 0 | 0 | 0 | **1** | 0 | 0 | 0 |
| **$|22\rangle$** | 0 | 0 | 0 | 0 | 0 | 0 | 0 | 0 | **1** |

**Figure 6.** The quantum circuit of our proposed non-phase relative ternary SWAP gate, with the total quantum cost ($QC_3$) of nine ternary gates (three one-qutrit gates and six two-qutrit gates). Notice that all CH, CH$^\dagger$, permutative, and controlled-shift gates in this quantum circuit are non-decomposed ternary gates.

**Table 2.** Stage-by-stage analysis of our proposed non-phase relative ternary SWAP gate.

| **Initial states** $|q_b\ q_a\rangle$ | **Entanglement stage** | **Un-entanglement stage** | **State correction stage** | **Phase correction stage** $|q_a\ q_b\rangle$ |
|---|---|---|---|---|
| $|0\ 0\rangle$ | $|0\ 0\rangle + |1\ 1\rangle + |2\ 2\rangle$ | $|0\ 0\rangle$ | $|0\ 0\rangle$ | $|0\ 0\rangle$ |
| $|0\ 1\rangle$ | $|0\ 0\rangle + \omega|1\ 1\rangle + \omega^2|2\ 2\rangle$ | $|1\ 0\rangle$ | $|1\ 0\rangle$ | $|1\ 0\rangle$ |
| $|0\ 2\rangle$ | $|0\ 0\rangle + \omega^2|1\ 1\rangle + \omega|2\ 2\rangle$ | $|2\ 0\rangle$ | $|2\ 0\rangle$ | $|2\ 0\rangle$ |
| $|1\ 0\rangle$ | $|1\ 0\rangle + |2\ 1\rangle + |0\ 2\rangle$ | $|0\ 2\rangle$ | $|0\ 1\rangle$ | $|0\ 1\rangle$ |
| $|1\ 1\rangle$ | $|1\ 0\rangle + \omega|2\ 1\rangle + \omega^2|0\ 2\rangle$ | $\omega^2|1\ 2\rangle$ | $\omega^2|1\ 1\rangle$ | $|1\ 1\rangle$ |
| $|1\ 2\rangle$ | $|1\ 0\rangle + \omega^2|2\ 1\rangle + \omega|0\ 2\rangle$ | $\omega|2\ 2\rangle$ | $\omega|2\ 1\rangle$ | $|2\ 1\rangle$ |
| $|2\ 0\rangle$ | $|2\ 0\rangle + |0\ 1\rangle + |1\ 2\rangle$ | $|0\ 1\rangle$ | $|0\ 2\rangle$ | $|0\ 2\rangle$ |
| $|2\ 1\rangle$ | $|2\ 0\rangle + \omega|0\ 1\rangle + \omega^2|1\ 2\rangle$ | $\omega|1\ 1\rangle$ | $\omega|1\ 2\rangle$ | $|1\ 2\rangle$ |
| $|2\ 2\rangle$ | $|2\ 0\rangle + \omega^2|0\ 1\rangle + \omega|1\ 2\rangle$ | $\omega^2|2\ 1\rangle$ | $\omega^2|2\ 2\rangle$ | $|2\ 2\rangle$ |

Moreover, a generic ternary Galois Field (GF3) circuit ($f$) of modulo-3 multiplication ($\cdot_3$) and modulo-3 addition ($+_3$) is cost-effectively designed for all possible states ($|0\rangle$, $|1\rangle$, and $|2\rangle$) of two input qutrits (the controls $a$ and $b$) and one output qutrit (the target $c$), as stated in Eq. (25) and shown in Fig. 7, where the ternary $\cdot_3$ and $+_3$ operators are expressed in Table 3 and Table 4, respectively. Notably, when $c = |0\rangle$, a cost-effective three-qutrit Toffoli gate, $f = (a \cdot_3 b)$, is generated. The quantum cost ($QC_3$) of this GF3 circuit is ten ternary gates of only two-qutrit permutative and shift gates. Industrially, the advantage of this generic GF3 circuit design is that all controls are only connected to the target, yielding to easily route the target between the controls for efficiently mapping the logical into physical qutrits in the restricted layout (square-grid and heavy-hex neighboring connectivity) of future ternary superconducting quantum computers.

$$f = (a \cdot_3 b) +_3 c;\ \forall\ a, b, c \in \{0, 1, 2\} \tag{25}$$

**Table 3.** The Marquand chart of the ternary Galois Field (GF3) multiplication ($\cdot_3$) operator.

| $\cdot_3$ | **0** | **1** | **2** |
|---|---|---|---|
| **0** | 0 | 0 | 0 |
| **1** | 0 | 1 | 2 |
| **2** | 0 | 2 | 1 |

**Table 4.** The Marquand chart of the ternary Galois Field (GF3) addition ($+_3$) operator.

| $+_3$ | **0** | **1** | **2** |
|---|---|---|---|
| **0** | 0 | 1 | 2 |
| **1** | 1 | 2 | 0 |
| **2** | 2 | 0 | 1 |

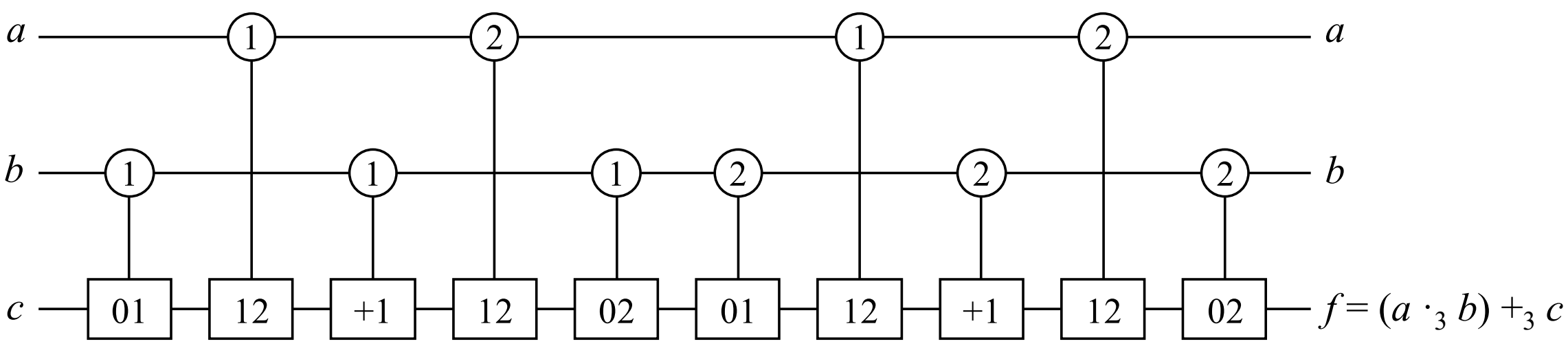

**Figure 7.** The quantum circuit of our generic cost-effective ternary Galois Field (GF3), $f = (a \cdot_3 b) +_3 c$, for all possible ternary states ($|0\rangle$, $|1\rangle$, and $|2\rangle$) of two controls ($a$ and $b$) and one target ($c$). The total quantum cost ($QC_3$) of $f$ is ten ternary gates of only two-qutrit permutative and shift (non-decomposed) gates. Notice that when $c = |0\rangle$, a cost-effective three-qutrit Toffoli gate is generated, $f = (a \cdot_3 b)$.

In conclusion, these non-phase relative two-qutrit SWAP gate, generic three-qutrit GF3 circuit, and cost-effective three-qutrit Toffoli gate are proposed here as an open quantum research topic (as a multi-qutrit circuit framework) of how to appropriately choose the ternary phase ($Z_3$ and $Z_3^\dagger$), superposition (TSGs, CH, and $CH^\dagger$), permutative, and shift gates (including all their corresponding controlled gates) for successfully operating all possible states ($|0\rangle$, $|1\rangle$, and $|2\rangle$) of multi-qutrit quantum circuits. Please observe that, in binary quantum computing, we introduced the "Bloch sphere approach (BSA)" [26–33] as an efficient geometrical design tool to visually construct important cost-effective binary $n$-qubit circuits, without utilizing any $2^n \times 2^n$ matrices multiplication ($n \geq 2$). Similarly, in ternary quantum computing, our future research will focus on developing an efficient ternary visualization structure (as a multi-dimensional quantum space representation) to geometrically construct cost-effective $m$-qutrit circuits, without utilizing any $3^m \times 3^m$ matrices multiplication ($m \geq 1$), as mathematically discussed along with toric geometrical presentation in our previous work [34].

## References

[1] A. Steane, "Quantum computing," *Reports on Progress in Physics*, vol. 61, no. 2, pp. 117–173, 1998.

[2] T. Hey, "Quantum computing: An introduction," *Computing and Control Engineering*, vol. 10, no. 3, pp. 105–112, 1999.

[3] J.L. O'brien, "Optical quantum computing," *Science*, vol. 318, no. 5856, pp. 1567–1570, 2007.

[4] M.A. Nielsen and I.L. Chuang, *Quantum Computation and Quantum Information*, 10th ed., Cambridge University Press, 2010.

[5] R. LaPierre, *Introduction to Quantum Computing*, 1st ed. Springer, 2021.

[6] L.K. Grover, "A fast quantum mechanical algorithm for database search," In *Proc. of the 28th Ann. ACM Symp. on Theory of Computing*, 1996, pp. 1212–219.

[7] G. Brassard, P. Høyer, and A. Tapp, "Quantum counting." In *Automata, Languages and Programming*, Springer, 1998.

[8] C. Figgatt, D. Maslov, K.A. Landsman, N.M. Linke, S. Debnath, and C. Monroe, "Complete 3-qubit Grover search on a programmable quantum computer," *Nature Communications*, vol. 8, no. 1, p. 1918, 2017.

[9] A. Al-Bayaty and M. Perkowski, "A concept of controlling Grover diffusion operator: A new approach to solve arbitrary Boolean-based problems," *Scientific Reports*, vol. 14, no. 1, p. 23570, 2024.

[10] A. Al-Bayaty and M. Perkowski, "BHT-QAOA: The generalization of quantum approximate optimization algorithm to solve arbitrary Boolean problems as Hamiltonians," *Entropy*, vol. 26, no. 10, p. 843, 2024.

[11] A. Al-Bayaty and M. Perkowski, "The impact of optimization approximation algorithms on the performance of the BHT-QAOA," *Academia Quantum*, vol. 2, no. 4, 2025.

[12] C. Moraga, "On some basic aspects of ternary reversible and quantum computing,". In *2014 IEEE 44th International Symposium on Multiple-Valued Logic*, May 2014, pp. 178–183.

[13] B. Cambou, P.G. Flikkema, J. Palmer, D. Telesca, and C. Philabaum, "Can ternary computing improve information assurance?," *Cryptography*, vol. 2, no. 1, p. 6, 2018.

[14] F.S. Khan and M. Perkowski, "Synthesis of ternary quantum logic circuits by decomposition," 2005, *arXiv:quant-ph/0511041*.

[15] Y. Wang and M. Perkowski, "Improved complexity of quantum oracles for ternary Grover algorithm for graph coloring," In *2011 41st IEEE International Symposium on Multiple-Valued Logic*, May 2011, pp. 294–301.

[16] G. Yang, X. Song, M. Perkowski, and J. Wu, "Realizing ternary quantum switching networks without ancilla bits," *Journal of Physics A: Mathematical and General*, vol. 38, no. 44, pp. 9689–9697, 2005.

[17] H.E. Chrestenson, "A class of generalized Walsh functions," *Pacific Journal of Mathematics*, vol. 5, pp. 17–31, 1955.

[18] S.B. Mandal, A. Chakrabarti, and S. Sur-Kolay, "Synthesis of ternary Grover's algorithm," In *2014 IEEE 44th International Symposium on Multiple-Valued Logic*, May 2014, pp. 184–189.

[19] M. Perkowski, "Quantum robots. Now or never?," *Invited Talk at the 5th National Conference on Informatics*, Gdansk, Poland, May 2007.

[20] A. Al-Rabadi, L. Casperson, M. Perkowski, and X. Song, "Multi-valued quantum logic," *Quantum*, vol. 10, no. 2, 2002.

[21] N. Mayo, T. Mor, and Y. Weinstein, "Benchmarking quantum computers via protocols, comparing IBM's Heron vs IBM's Eagle," 2026, *arXiv:2603.04377*.

[22] S. Shirgure, E. Kökcü, A. Mitra, W.A. de Jong, C. Iancu, and S. Niu, "Characterizing and benchmarking dynamic quantum circuits," 2026, *arXiv:2604.03360*.

[23] F. Pudda, M. Chizzini, and L. Crippa, "Generalised quantum gates for qudits and their application in quantum Fourier transform," 2024, *arXiv:2410.05122*.

[24] N. Goss, A. Morvan, B. Marinelli, B.K. Mitchell, L.B. Nguyen, R.K. Naik, L. Chen, C. Jünger, J.M. Kreikebaum, D.I. Santiago, and J.J. Wallman, "High-fidelity qutrit entangling gates for superconducting circuits," *Nature Communications*, vol. 13, no. 1, p. 7481, 2022.

[25] A. Al-Bayaty, A. Al-Shuwaili, A. AlZubaidi, and M. Perkowski, "Improving the circuit realization of Grover's quantum search algorithm by replacing Hadamard with √X gates," In *Proceedings of the 2026 2nd International Conference on Computing and Emerging Sciences (ICCES '26)*, Association for Computing Machinery, New York, NY, USA, Feb. 2026, pp. 258–265.

[26] A. Al-Bayaty and M. Perkowski, "p-SWAP: A generic cost-effective quantum Boolean-phase SWAP gate using two CNOT gates and the Bloch sphere approach," 2024, *arXiv:2410.16641*.

[27] A. Al-Bayaty, S. Chen, S.A. Bleiler, and M. Perkowski, "A cost-effective quantum Boolean-phase SWAP gate with only two CNOT gates," 2025, *arXiv:2507.17164*.

[28] A. Al-Bayaty and M. Perkowski, "A geometrical design tool for building cost-effective layout-aware n-bit quantum gates using the Bloch sphere approach," 2026, *arXiv:2601.00484*.

[29] A. Al-Bayaty and M. Perkowski, "BSA: The Bloch sphere approach as a geometrical design tool for building cost-Effective quantum gates," 2024, *Protocols.io*.

[30] A. Al-Bayaty and M. Perkowski, "Cost-effective realization of n-bit Toffoli gates for IBM quantum computers using the Bloch sphere approach and IBM native gates," 2024, *arXiv:2410.13104*.

[31] A. Al-Bayaty and M. Perkowski, "GALA-n: Generic architecture of layout-aware n-bit quantum operators for cost-effective realization on IBM quantum computers," 2023, *arXiv:2311.06760*.

[32] A. Al-Bayaty, X. Song, and M. Perkowski, "CALA-n: A quantum library for realizing cost-effective 2-, 3-, 4-, and 5-bit gates on IBM quantum computers using Bloch sphere approach, Clifford+T gates, and layouts," 2024, *arXiv:2408.01025*.

[33] A. Al-Bayaty, "Layout-aware quantum circuitry and algorithmic extensions to Grover's algorithm," Ph.D. dissertation, Portland State University, Portland, OR, USA, 2025.

[34] S. Bleiler, A. Al-Bayaty, S. Chen, and M. Perkowski, "Visualizing the state space of quantum trits, quadits, and pairs of qubits via toric geometry," 2025, *arXiv:2510.01455*.